\documentclass[final,3p,times]{elsarticle}
\usepackage{amssymb}
\usepackage{amsmath}
\journal{Physica A}

\def\Eq#1{Eq.~(\ref{#1})}
\def\Eqs#1{Eqs.~(\ref{#1})}

\def\no{\nonumber\\}

\def\>{\rangle}
\def\<{\langle}

\def\half{\frac{1}{2}}
\def\adg{a^\dagger}
\def\adgsq{a^{\dagger\, 2}}
\def\tr{\text{tr}}
\def\dg{\dagger}
\def\Kt{\bar{K}}
\def\rhot{\bar{\rho}}
\def\sigt{\bar{\sigma}}
\def\Delt{\bar{\Delta}}
\def\xt{\bar{x}}
\def\pt{\bar{p}}
\def\etat{\bar{\eta}}
\def\lam{\lambda}

\def\w{\omega}

\def\dxx{\sigma_{\!xx}}
\def\dpp{\sigma_{\!pp}}
\def\dxp{\sigma_{\!xp}}
\def\Del{\Delta}
\def\gam{\gamma}

\def\al{\alpha}

\def\th{\theta}

\def\d{\partial}
\def\KL{\text{KL}}
\def\CL{\text{CL}}
\def\HPZ{\text{HPZ}}
\def\dt{\text{d}}
\def\T{\text{T}}
\def\etav{\mbox{\boldmath$\eta$}}
\def\etatv{\bar{\mbox{\boldmath$\eta$}}}
\def\Iv{\mbox{\boldmath$I$}}
\def\uv{\mbox{\boldmath$u$}}
\def\vv{\mbox{\boldmath$v$}}
\def\Kv{\mbox{\boldmath$K$}}
\def\Uv{\mbox{\boldmath$U$}}
\def\Vv{\mbox{\boldmath$V$}}
\def\tth{\hat{\theta}}
\def\phih{\hat{\phi}}
\def\psih{\hat{\psi}}
\def\gamh{\hat{\gamma}}
\def\dh{\hat{d}}

\begin{document}

\begin{frontmatter}

\title{Damping modes of harmonic oscillator in open quantum systems}

\author[1]{B. A. Tay}
\ead{BuangAnn.Tay@nottingham.edu.my}
\address[1]{Department of Foundation in Engineering, Faculty of Science and Engineering, University of Nottingham Malaysia Campus, Jalan Broga, 43500 Semenyih, Selangor, Malaysia}

\date{\today}

\begin{abstract}
Through a set of generators that preserves the hermiticity and trace of density matrices, we analyze the damping of harmonic oscillator in open quantum systems into four modes, distinguished by their specific effects on the covariance matrix of position and momentum of the oscillator. The damping modes could either cause exponential decay to the initial covariance matrix or shift its components. They have to act together properly in actual dynamics to ensure that the generalized uncertainty relation is satisfied. We use a few quantum master equations to illustrate the results.

\end{abstract}


\end{frontmatter}

\section{Introduction}

The time evolution of an isolated quantum oscillator is generated by its free Hamiltonian. The unitary time evolution operator can be interpreted as a rotation operator because it causes the center of the oscillator's wave packet to trace a circular path in the phase space \cite{Hillery84,Balazs84}. Two other familiar unitary operators are the displacement operator \cite{Klauder68,Gardiner} that shifts the center of the wave packet in the phase space, and the squeezed operator \cite{Walls83,Gardiner} that turns the circular profile of a wave packet into an ellipse. The effects of these operators on the quantum state of oscillator are revealed by the change in the profile of the oscillator's wave packet, which is characterized by the mean and covariance matrix of the phase space coordinates.

When we consider open quantum systems, dissipative operators get involved. For example, under the Kossakowski-Lindblad (KL) equation \cite{Kossa76,Lindblad76}, the oscillator undergoes amplitude damping \cite{Nielsen}. The Caldeira-Leggett (CL) equation \cite{CL83} and the Hu-Paz-Zhang (HPZ) equation \cite{HPZ92} cause the quantum oscillator to relax as it evolves in a thermal bath and affect the quantum correlation it carries. Though there had been extensive studies in the mechanism of damping and decoherence, see Refs.~\cite{Gardiner,Weiss,Breuer,Decoh03} and references therein, a separation of dissipation into the specific ways or modes in which damping could occur are not yet carried out, partly due to the difficulty in analyzing the dissipative part of the generators of time evolution into suitable components. Examples of damping modes are provided by the relaxation of two-level systems or qubits. The damping in qubits \cite{Ruskai02,Zyczkowski04,Tay19} can be analyzed by following the change in the Bloch ball \cite{Nielsen}. Unitary rotations generated by the Pauli matrices change the axis of the Bloch ball. In dissipative environment, damping distorts the Bloch ball into an ellipsoid, or shifts the center of the Bloch ball. In this work we will analyze the damping of the quantum states of harmonic oscillator in a similar spirit.

It is challenging to analyze the dissipative part of the generator of time evolution in terms of meaningful damping modes partly because dissipation turns the pure state of oscillator into mixed state. Mathematically this implies that the operators are non-distributive \cite{Prigogine73,Sungyun03}, and hence cannot be written as a simple product of unitary operators $u\times u^\dg$. There was effort to set up a basis of operators in terms of commutator brackets \cite{Tameshtit13}. Recently, we construct a different basis of seven generators that are bilinear in the creation and annihilation operators of harmonic oscillator \cite{Tay17}. They generate transformations that preserve the hermiticity and the trace of density matrices. Three of them correspond to the rotation and squeezed operators. When we analyze generic damping in terms of the other four dissipative operators, we obtain simple picture in which the various modes would change the covariance matrix of the phase space coordinates of quantum states in specific ways. It is the main purpose of this work to clarify the effects of these damping modes on the quantum states of harmonic oscillator.

\section{Moments and unitary operators}
\label{SecMomUniOp}

\subsection{Moments of transformed states}

The dimensionless position and momentum operators of harmonic oscillator are $x=(a+\adg)/\sqrt{2}$ and $p=i(\adg-a)/\sqrt{2}$, respectively, where $\adg$ and $a$ are creation and annihilation operators, and we use the unit $\hbar=1$. The moments of these operators can be used to characterize the properties of quantum state described by the density matrix $\rho$. In particular, Gaussian wave packets are fully characterized by the first order moments (mean) and the second order moments (covariance matrix or uncertainty) of the position and momentum operators. Hence, to study the effects of damping, we follow the change in the following quantities,
\begin{subequations}
\begin{align} \label{def}
    \<x\>&=\frac{1}{\sqrt{2}}\big(\<a\>+\<\adg\>\big)\,,&\qquad
    \<p\>&=\frac{i}{\sqrt{2}}\big(\<\adg\>-\<a\>\big)\,,\\
    \dxx&\equiv\<x^2\>-\<x\>^2\,,&\qquad
    \dpp&\equiv\<p^2\>-\<p\>^2\,,&\qquad
    \dxp&\equiv\half \<xp+px\>-\<x\>\<p\>\,,\label{dxxdef}\\
    \Del&\equiv\dxx\dpp-\dxp^2\,,\label{Deldef}
\end{align}
\end{subequations}
where the expectation value of an operator $o$ is given by $\<o\>\equiv \tr(o\rho)$.
The last quantity $\Del$ is the generalized uncertainty \cite{Dekker84} of quantum state. Physical states must satisfy the generalized uncertainty relation $\Del\geq \frac{1}{4}$. The coherent and squeezed states are minimum uncertainty states that minimize the generalized uncertainty relation. Gaussian states satisfy the generalized uncertainty relation if and only if they have positive time evolution \cite{Tay17,Talkner81}.

To study the effect of an operator $S$ that transformed the state into $\rho'\equiv S\rho$, we find out how the expectation value of an operator $\<o\>'\equiv \tr(o\rho')$ is modified. This quantity can be evaluated by passing $S$ over to act on the operator
\begin{align}   \label{o'}
    \<o\>'=\tr(o\cdot S\rho)=\tr(S^\T o\cdot \rho)=\<o'\>\,,
\end{align}
where the transformed operator is
\begin{align}   \label{o'def}
    o'\equiv S^\T o\,.
\end{align}
$S^\T$ denotes the transposition operation on $S$ \cite{Prigogine73,Tay17} that can be obtained by using the invariance of trace under cyclic permutations of operators. We restrict our consideration to $S$ that can be written as a sum of superoperators $a_i\times b_i$, that is, $S=\sum_i \mu_i (a_i\times b_i)$, where $\mu_i$ are complex numbers. Superoperator acts on operator as $(a_i\times b_i) \cdot \rho=a_i\rho b_i$. The invariance of trace under the cyclic permutations of operators then enables us to obtain
\begin{align}   \label{STA}
    \<o\>'\equiv \sum_i \mu_i \tr(oa_i\rho b_i)=\sum_i \mu_i \tr(b_i o a_i\cdot\rho )= \tr\bigg(\sum_i \mu_i \big(b_i\times a_i\big) o\cdot\rho \bigg)\,.
\end{align}
Comparing the right hand side of \Eq{STA} with the right hand side of \Eq{o'}, we deduce that
\begin{align}   \label{Sadj}
    S^\T=\sum_i \mu_i (b_i\times a_i) \,.
\end{align}

\subsection{Unitary operators}
\label{SecUni}

We restrict our consideration to generators that are bilinear in the annihilation and creation operators. From the results obtained in Ref.~\cite{Tay17}, there are seven generators that give rise to transformation that preserve the hermiticity and trace of density matrices. Three of the generators among them \cite{Tay17},
\begin{subequations}
\begin{align} \label{iL0A}
    iL_0&\equiv\frac{i}{2}(\adg a\times 1-1\times \adg a)\,, \\
    iM_1&\equiv\frac{i}{4}(\adgsq\times 1+a^2\times 1-1\times \adgsq-1\times a^2)\,,\\
    iM_2&\equiv\frac{1}{4}(\adgsq\times 1-a^2\times 1-1\times \adgsq+1\times a^2)\,,
\end{align}
\end{subequations}
generate the familiar unitary operators.

The first one $iL_0\rho=i[\adg a,\rho]$ generates the time evolution of free oscillator.
We work out $a'=(e^{\th iL_0})^\T a$ as an illustration of how we calculate the transformed operators.
We can simplify the action of the operator by using the commutation relation $[a,\adg]=a\adg-\adg a=1$. For example, $L_0a=\frac{1}{2}(\adg aa-a \adg a)=-\half a$ and $L_0\adg=\half\adg$. Applying these equations repeatedly, and using the fact that $L_0$ is antisymmetric under transposition, $(iL_0)^\T=\frac{i}{2}(1\times \adg a-\adg a\times 1)=-iL_0$, we obtain
\begin{align} \label{expiL0}
    a'= \big(e^{\th iL_0}\big)^\T a&=e^{-\th iL_0}a
    =\sum_{n=0}^\infty \frac{(-i\th)^n}{n!} L_0^n a
    =\sum_{n=0}^\infty \frac{(i\th/2)^n}{n!} a
    =e^{i\th/2} a\,.
\end{align}
Together with the hermitian conjugate of $a'$, we obtain the mean in the transformed position $\<x'\>$ and momentum $\<p'\>$ operators from \Eq{def}, listed in the second column of Table \ref{tab1}. We see that $e^{\th iL_0}$ causes an ordinary rotation in the phase space coordinates. Similarly, we can calculate the transformed bilinear operators $(aa)', (\adg a)'$ and $(\adg\adg)'$ to obtain the covariance matrix of position and momentum. \Eqs{dxxdef} and \eqref{Deldef} then give the rests of the results listed in the second column of Table \ref{tab1}.

The generators $iM_1$ and $iM_2$ are related to the squeezed operator $s=\exp(-\frac{1}{2}\xi\adgsq+\frac{1}{2}\xi^*a^2)$ \cite{Walls83,Gardiner} with complex squeezing parameter $\xi=\xi_\text{R}+i\xi_\text{I}$ as follows. In the density matrix space, squeezed operator acts on density matrix as $(s\times s^\dg)\rho=s\rho s^\dg$. Denoting $s_R=s|_{\xi=\xi_R}$ and $s_I=s|_{\xi=\xi_I}$, $iM_2$ and $iM_1$ are related to the real and imaginary part of the squeezing parameter by $s_R\times s^\dg_R=\exp(-2\xi_R iM_2)$ and $s_I\times s^\dg_I=\exp(-2\xi_I iM_1)$, respectively.

The mean and covariance matrix of the transformed position and momentum can be calculated similarly. The results are listed in the third and fourth column of Table \ref{tab1}. We find that $iM_1$ generates a hyperbolic rotation in the phase space coordinates, whereas $iM_2$ expands one of the coordinates at the expense of a contraction in the other. As a result, even though the variances of the position and momentum are changed, the overall generalized uncertainty remains the same, $\Del'=\Del$. This is consistent with the unitary nature of squeezed operator.

\begin{table}[t]
\center
\begin{tabular}{cccccccc}
    \hline\hline\\[-1.5ex]
  $S$ &\quad &$e^{\th iL_0}$ & \quad & $e^{\phi iM_1}$ & \quad   & $e^{\psi iM_2}$ \vspace{3pt}\\
  \hline\\[-1.5ex]
  $S^\T$ &\quad &$e^{-\th iL_0}$ & \quad & $e^{-\phi iM_1}$ & \quad   & $e^{-\psi iM_2}$
  \vspace{3pt}\\
  \\[-1.5ex]
  $\<1'\>$ &\quad &1 &\quad &1 &\quad & 1  \vspace{3pt}\\
  \\[-1.5ex]
    $\<x'\>$ &\quad & $\cos\frac{\th}{2}\cdot\<x\>-\sin\frac{\th}{2}\cdot\<p\>$ &\quad &$\cosh\frac{\phi}{2}\cdot\<x\>+\sinh\frac{\phi}{2}\cdot\<p\>$ &\quad &$e^{\psi/2}\<x\>$\vspace{6pt}\\
    $\<p'\>$ &\quad & $\sin\frac{\th}{2}\cdot\<x\>+\cos\frac{\th}{2}\cdot\<p\>$ &\quad &$\sinh\frac{\phi}{2}\cdot\<x\>+\cosh\frac{\phi}{2}\cdot\<p\>$ &\quad
  &$e^{-\psi/2}\<p\>$\vspace{6pt}\\
    \\[-1.5ex]
  $\dxx'$    &\quad & $\cos^2\frac{\th}{2}\cdot\dxx +\sin^2\frac{\th}{2}\cdot\dpp-\sin\th \cdot\dxp $
            & \quad & $\cosh^2\frac{\phi}{2}\cdot\dxx+\sinh^2\frac{\phi}{2}\cdot\dpp+\sinh\phi\cdot\dxp$
            & \quad & $e^{\psi}\dxx$
             \vspace{6pt}\\
  $\dpp'$ &\quad & $\sin^2\frac{\th}{2}\cdot\dxx +\cos^2\frac{\th}{2}\cdot\dpp+\sin\th\cdot\dxp  $
        & \quad & $\sinh^2\frac{\phi}{2}\cdot\dxx+\cosh^2\frac{\phi}{2}\cdot\dpp+\sinh\phi\cdot\dxp$
        & \quad & $e^{-\psi}\dpp$ \vspace{6pt}\\
  $\dxp'$ &\quad & $\sin\frac{\th}{2}\cos\frac{\th}{2}\cdot(\dxx-\dpp)  +\cos\th\cdot\dxp$
        & \quad & $\sinh\frac{\phi}{2}\cosh\frac{\phi}{2}\cdot(\dxx+\dpp)  +\cosh\phi\cdot\dxp$
        & \quad & $\dxp$
        \vspace{6pt}\\
  \\[-1.5ex]
  $\Del'$ &\quad & $\Del$ & \quad & $\Del$ & \quad & $\Del$ \\
  \hline\hline\\
\end{tabular}
\caption{Moments and generalized uncertainty of quantum state under the action of unitary operators.}
\label{tab1}
\end{table}

\section{Damping modes}
\label{SecDampMod}

\subsection{Dissipative operators}
\label{SecDissop}

From the results of Ref.~\cite{Tay17}, there are four generators that give rise to damping. The four dissipative generators are
\begin{subequations}
\begin{align}
    O_0-I/2&\equiv\frac{1}{2}(\adg\times a-a\times \adg-1\times 1) \,,\\
    O_+&\equiv\frac{1}{2}(\adg\times a+a\times \adg-\adg a\times 1-1\times \adg a-1\times 1)\,,\\
    L_{1+}&\equiv\frac{1}{4}(2\adg \times \adg+2a\times a-\adgsq\times 1-a^2\times 1-1\times \adgsq-1\times a^2)\,,\\
    L_{2+}&\equiv-\frac{i}{4}(2\adg \times \adg-2a\times a-\adgsq\times 1+a^2\times 1-1\times \adgsq +1\times a^2 )\,.
\end{align}
\end{subequations}
The mean and covariance matrix of the transformed position and momentum under the actions of the dissipative operators can be calculated straight-forwardly. The results are listed in Table \ref{tab2}.

\begin{table}[t]
\center
\begin{tabular}{ccccccccc}
\hline\hline\\[-1.5ex]
  $S$ &\quad &$e^{\al (O_0-I/2)}$ &\quad &$e^{\eta_0 O_+}$ &\quad &$e^{\eta_1 L_{1+}}$ &\quad &$e^{\eta_2 L_{2+}}$
  \vspace{3pt}\\
  \hline\\[-1.5ex]
  $S^\T$ &\quad &$e^{-\al (O_0+I/2)}$ &\quad &$e^{\eta_0 O_+}$ &\quad &$e^{\eta_1 L_{1+}}$ &\quad &$e^{\eta_2 L_{2+}}$ \vspace{3pt}\\
  \\[-1.5ex]
  $\<1'\>$ &\quad &1 &\quad &1 &\quad & 1 &\quad &1 \vspace{3pt}\\
  \\[-1.5ex]
  $\<x'\>$ &\quad & $e^{\al/2}\<x\>$ &\quad & $\<x\>$ & \quad &$\<x\>$ &\quad &$\<x\>$ \vspace{6pt}\\
  $\<p'\>$ &\quad & $e^{\al/2}\<p\>$ &\quad & $\<p\>$ & \quad &$\<p\>$ &\quad &$\<p\>$ \vspace{6pt}\\
  \\[-1.5ex]
  $\dxx'$ &\quad & $e^\al\dxx $ &\quad & $\dxx +\frac{\eta_0}{2}$ & \quad &$\dxx -\frac{\eta_1}{2}$ &\quad &$\dxx$ \vspace{6pt}\\
  $\dpp'$ &\quad & $e^\al\dpp $ &\quad & $\dpp +\frac{\eta_0}{2}$ & \quad &$\dpp +\frac{\eta_1}{2}$ &\quad &$\dpp$ \vspace{6pt}\\
  $\dxp'$ &\quad & $e^\al\dxp $ &\quad & $\dxp$ & \quad &$\dxp$ &\quad &$\dxp +\frac{\eta_2}{2}$ \vspace{6pt}\\
  \\[-1.5ex]
  $\Del'$ &\quad & $e^{2\al}\Del $ &\quad & $\Del+\frac{\eta_0}{2}(\dxx+\dpp)+\frac{\eta_0^2}{4}$ & \quad &$\Del+\frac{\eta_1}{2}(\dxx-\dpp)-\frac{\eta_1^2}{4}$ &\quad &$\Del-\eta_2 \dxp-\frac{\eta_2^2}{4}$ \vspace{3pt}\\
  \hline\hline\\
\end{tabular}
\caption{Moments and generalized uncertainty of quantum state under the action of dissipative operators.}
\label{tab2}
\end{table}

We find that $O_0-I/2$ causes expansion in both of the mean as well as the covariance matrix. On the contrary, $O_+, L_{1+}$ and $L_{2+}$ keep the mean invariant, while alter the covariance matrix in complimentary ways. While $O_+$ and $L_{1+}$ shift the variances $\dxx, \dpp$, it leaves the covariance $\dxp$ intact. In view of this, the effects of $O_+$ and $L_{1+}$ are similar to their classical counterparts in the kinetic equation that would cause damping and diffusion to the oscillator's wave packet.

On the other hand, $L_{2+}$ acts in a non-classical way. It leaves the variances unchanged, but shifts the covariance. Because of this, $L_{2+}$ is sometimes referred to as the ``anomalous diffusion" \cite{Barsegov02} when it appears in the HPZ equation \eqref{KHPZ}. For the convenience of the readers, the representation of the operators in the phase space coordinates are listed in the appendix, cf. \Eqs{iL0xp}-\eqref{L2+xp}. Notice that $L_{2+}$ is $\frac{1}{2}\d^2/\d p \d q$ \eqref{L2+xp} in the phase space coordinates, a cross term that seldom appears in actual classical kinetic equations \cite{Risken}.

Notice that the generalized uncertainty must change under the dissipative operators. In contrast to the unitary operators that impose no restriction on their parameters, the range of the parameters of dissipative operators must be restricted to satisfy the generalized uncertainty relation $\Del' \geq \frac{1}{4}$.

From the above analysis we learn that there are four damping modes of harmonic oscillator in open quantum systems. The modes do not appear isolatedly in the dynamics, but must couple with each other in a correct proportion to satisfy the generalized uncertainty relation. This is manifested in the master equations considered in Sec.~\ref{SecRedDyn}.

\subsection{Evolution of damped oscillator}
\label{SecEvoOsc}

In illustration of the effects of damping modes, let us consider the evolution of damped oscillator governed by the master equation $\partial\rho/\partial t=-(K_0+K_\dt)\rho$. The unitary and dissipative part of the dynamics have the generic form \cite{Tay17},
\begin{align}   \label{K0}
    K_0&\equiv \th iL_0+\phi iM_1+\psi iM_2\,,\\
    K_\dt&\equiv\gam(O_0-I/2)+\eta_0 O_++\eta_1 L_{1+}+\eta_2 L_{2+}\,, \label{Kd}
\end{align}
respectively, where all the coefficients are real. $\gam$ is the relaxation constant of the system, and $\eta_i$ induces damping and diffusion to the wave packet of the oscillator.

Since we are familiar with the effects of the unitary part, we focus on the effects of the dissipative part by switching to the interaction picture. We denote quantities in the interaction picture by bars. The state in the interaction picture is $\rhot(t)\equiv e^{K_0t}\rho(0)$. It evolves as $\partial\rhot/\partial t=-\Kt_\dt(t)\rhot$ with the generator
\begin{align}   \label{Kt}
        \Kt_\dt(t)&\equiv e^{K_0 t}K_\dt e^{-K_0 t}
            =K_\dt +t[K_0,K_\dt]+\frac{t^2}{2!}[K_0,[K_0,K_\dt]]+\cdots\,.
\end{align}
In \ref{AppKd}, we show that
\begin{align}   \label{Kdt}
        \Kt_\dt(t)&=\gam(O_0-I/2)+\etat_0(t) O_++\etat_1(t) L_{1+}+\etat_2(t) L_{2+} \,.
\end{align}
The relaxation constant $\gam$ is not affected by a change to interaction picture. However, the other coefficients $\etat_i(t)$ are modified. It is further shown in \ref{AppKd} that they are given by the components of a vector
\begin{align}   \label{etatv}
        \etatv(t)
        &=\left(\uv_0\otimes \vv_0^\dg +\left(\Iv-\uv_0\otimes \vv_0^\dg\right)\cos \w t+\Kv_0 \sin\w t\right)\cdot \etav\,,
\end{align}
where
\begin{align}   \label{etav}
    \etatv(t)&\equiv \left(\begin{array}{c}
                             \etat_0(t) \\
                             \etat_1(t) \\
                             \etat_2(t)
                           \end{array}\right)\,, &\qquad
    \etav&\equiv \left(\begin{array}{c}
                             \eta_0 \\
                             \eta_1 \\
                             \eta_2
                           \end{array}\right)\,,
\end{align}
\begin{align}    \label{uv}
        \uv_0&= -i\left(\begin{array}{c}
                             \tth \\
                             \phih \\
                             \psih
                           \end{array}\right)\,,&\qquad
         \vv_0&= i\left(\begin{array}{c}
                             -\tth \\
                             \phih \\
                             \psih
                           \end{array}\right)\,,
\end{align}
\begin{align}   \label{K0v}
         \Kv_0\equiv \left(\begin{array}{ccc}
                             0&-\psih &\phih \\
                             -\psih &0 &\tth \\
                             \phih &-\tth &0
                           \end{array}\right)\,,
\end{align}
$\tth\equiv\th/\w, \phih\equiv\phi/\w, \psih\equiv \psi/\w$, in which
\begin{align}   \label{w}
         \w\equiv\sqrt{\th^2-\phi^2-\psi^2}
\end{align}
is assumed to be real, $\Iv$ is the $3\times 3$ identity matrix, $\dagger$ denotes hermitian conjugate and $\otimes$ denotes tensor product. $\uv_0$ and $\vv_0$ are the zero right and left eigenvectors of $\Kv_0$, respectively, cf. \ref{AppKd}.

The solution to the master equation in the interaction picture is $\rhot(t)=\left\{e^{-\int_{0}^{t}\Kt_\dt(t') dt'}\right\}_+\rhot(0)$, where $\{\cdot\}_+$ denotes time-ordered products. To solve for $\rhot(t)$, we decompose the time evolution operator into the following form \cite{Tay17b}
\begin{align}   \label{TmEvodecomp}
    \left\{e^{-\int_{0}^{t}\Kt_\dt(t') dt'}\right\}_+=e^{g_2(t)L_{2+}}e^{g_1(t)L_{1+}}e^{g_0(t)O_+}e^{h(t)(O_0-I/2)}\,,
\end{align}
where $h(t)$ and $g_i(0)$ are real coefficients. From the results of Ref.~\cite{Tay17b}, the coefficients satisfy the rate equations,
\begin{align}   \label{hg}
    \frac{dh}{dt}&=-\gam\,,&\qquad
    \frac{dg_i}{dt}&=-\etat_i-\gam g_i\,.
\end{align}
Using the initial condition $\left\{e^{-\int_{0}^{t}\Kt_\dt(t') dt'}\right\}_+\bigg|_{t=0}=I$, where $I$ is the identity operator, the coefficients have the initial conditions $h(0)=0=g_i(0)$. \Eq{hg} can now be solved to give
\begin{align}   \label{htgt}
    h(t)&=-\gam t\,,&\qquad
    g_i(t)&=-e^{\gam t}\int_0^t \etat_i(t')e^{-\gam t'}dt'\,.
\end{align}

Using $(AB)^\T=B^\T A^\T$ and the fact that $O_0$ is anti-symmetric under transposition, while $O_+, L_{1+}, L_{2+}$ are symmetric, we obtain
\begin{align}   \label{ot'}
    \left(\left\{e^{-\int_{0}^{t}\Kt_\dt(t') dt'}\right\}_+\right)^\T
        =e^{-h(t)(O_0+I/2)}e^{g_0(t)O_+}e^{g_1(t)L_{1+}}e^{g_2(t)L_{2+}}\,.
\end{align}
We then act the series of operators in \Eq{ot'} subsequently on the position and momentum operators using the results of Table \ref{tab2} to yield the following mean and covariance matrix of the transformed position and momentum,
\begin{subequations}
\begin{align}   \label{x'}
    \<\xt'\>&=e^{-\gam t/2}\<\xt\>\,,&\qquad
    \<\pt'\>&=e^{-\gam t/2}\<\pt\>\,,&\\
    \sigt'_{\!xx}&=e^{-\gam t}\sigt_{\!xx} +\half (g_0-g_1)\,,&\qquad
    \sigt'_{\!pp}&=e^{-\gam t}\sigt_{\!pp} +\half (g_0+g_1)\,,&\qquad
    \sigt'_{\!xp}&=e^{-\gam t}\sigt_{\!xp} +\half g_2\,,
\end{align}
\begin{align}
    \Delt'&=e^{-2\gam t}\Delt+\frac{1}{2} e^{-\gam t} \left[(g_0+g_1)\sigt_{\!xx}+(g_0-g_1)\sigt_{\!pp}-2g_2\sigt_{\!xp}\right]
    +\frac{1}{4}\big(g_{0}^2-g_{1}^2-g_{2}^2\big)\,,
\end{align}
\end{subequations}
where we have omitted the time dependence on the expressions for simplicity. We learn that $O_0-I/2$ cause the initial mean and covariance matrix to decay exponentially. The coefficients of $O_0-I/2$ in the master equation \eqref{Kd} therefore set the time scale of relaxation. Then specific components of the covariance matrix are shifted by the other three generators. In the long time limit $t\rightarrow \infty$, the generalized uncertainty goes into
\begin{align}   \label{Deltinf}
    \Delt'\rightarrow \frac{1}{4}(g_{0,\infty}^2-g_{1,\infty}^2-g_{2,\infty}^2)\,,
\end{align}
where $g_{i,\infty}$ denotes $g_i(t)$ in the limit $t\rightarrow \infty$. The quantity must satisfy the generalized uncertainty relation.
\Eq{Deltinf} suggests that $O_+$ tends to increase the generalized uncertainty of the state, whereas $L_{1+}$ and $L_{2+}$ counteract its effect. We observe that since $O_0-I/2$ eliminates the initial $\Del$ exponentially in the long time limit, it must be accompanied by at least $O_+$ if the master equation were to satisfy the generalized uncertainty relation. Moreover, $L_{1+}$ or $L_{2+}$ cannot appear in the master equation unless accompanied by $O_+$. The observations are indeed validated by the structure of the dissipative parts of the master equations discussed in the next section, see \Eqs{KKL}-\eqref{KHPZ}.

\subsection{Examples of reduced dynamics}
\label{SecRedDyn}

The generators of the Kossakowski-Lindblad (KL) equation \cite{Kossa76,Lindblad76}, the Caldeira-Leggett (CL) equation \cite{CL83} and the Hu-Paz-Zhang (HPZ) equation \cite{HPZ92} are
\begin{align}   \label{KKL}
    K_\KL&\equiv 2\w_0 iL_0+\gam (O_0-I/2)-2\gam b O_+\,,\\
    K_\CL&\equiv 2\w_0 iL_0+\gam(O_0-I/2-iM_2)-2\gam b(O_++L_{1+})\,,\label{KCL}\\
    K_\HPZ&\equiv 2\w_0 iL_0+\gam(O_0-I/2-iM_2)-2\gam b(O_++L_{1+})-dL_{2+}\,,\label{KHPZ}
\end{align}
respectively, where all the coefficients are real. $b$ is related to the temperature of the environment. $d$ produces the ``anomalous diffusion" already mentioned in Sec. \ref{SecDissop}.

Using the expressions of $\etat_i(t)$ obtained at the end of \ref{AppKd}, we find that for the KL equation,
\begin{align} \label{KLg}
       g_{0,\KL}(t)&=2b(1-e^{-\gam t})\,, &\qquad g_{1,\KL}(t)&=0=g_{2,\KL}(t)\,,
\end{align}
for the CL equation,
\begin{subequations}
\begin{align} \label{CLg}
       g_{0,\CL}(t)&=2b(1+\gamh^2-e^{-\gam t}-\gamh^2\cos \w t)\,, \\
       g_{1,\CL}(t)&=2b\gamh \sin \w t\,,\\
       g_{2,\CL}(t)&=2b\gamh\sqrt{1+\gamh^2}(\cos\w t-1)\,,
\end{align}
\end{subequations}
where $\w\equiv \sqrt{4\w_0^2-\gam^2}, \gamh\equiv \gam/\w$,
and for the HPZ equation,
\begin{subequations}
\begin{align} \label{HPZg}
       g_{0,\HPZ}(t)&=g_{0,\CL}(t)+\frac{\dh}{\sqrt{1+\gamh^2}}
                    (1+\gamh^2-e^{-\gam t}-\gamh^2\cos\w t-\gamh\sin\w t)\,, \\
       g_{1,\HPZ}(t)&=g_{1,\CL}(t)+\frac{\dh}{\sqrt{1+\gamh^2}} (e^{-\gam t}-\cos\w t+\gamh \sin \w t)\,,\\
       g_{2,\HPZ}(t)&=g_{2,\CL}(t)+\dh(-\gamh+\gamh\cos\w t+\sin \w t)\,,
\end{align}
\end{subequations}
in which $\dh\equiv d/\w$.

In the long time limit where the rapidly oscillating terms average to zero, the generalized uncertainty under the time evolution of the master equations can be approximated by
\begin{align}   \label{DelKL}
        \Delt'_\KL&\rightarrow b^2\,,\\
        \Delt'_\CL&\rightarrow b^2(1+\gamh^2)\,,\\
        \Delt'_\HPZ&\rightarrow (b+d/(4\w_0))^2\,. \label{DelHPZ}
\end{align}
This is also the generalized uncertainty in the Schr\"odinger picture since from Table \ref{tab1} we know that the unitary part of the dynamics does not affect the generalized uncertainty. From \Eq{DelHPZ} we notice that a negative $d$ comparable to $b$ has the danger of violating the generalized uncertainty relation. This is consistent with the result obtained in Ref.~\cite{Tay17b}.
When we subject \Eqs{DelKL}-\eqref{DelHPZ} to the condition $\Delt'\geq \frac{1}{4}$, they produce positivity conditions similar to those obtained from the results in Ref.~\cite{Tay17b} up to terms linear in $\gam$ and $d$.

\section{Conclusion}
\label{SecConclusion}

We analyze the damping of harmonic oscillator in open quantum systems into four independent modes through a set of dissipative operators. These modes do not occur isolatedly in the actual time evolution of the oscillator. They must act alongside with others properly to produce time evolution that satisfies the generalized uncertainty relation.
Among the generators, $O_0-I/2$ sets the time scale of relaxation and eliminates the initial mean and covariance matrix. $O_+$ and $L_{1+}$ causes diffusion to the wave packets in the phase space coordinates. As a result, the variances of position and momentum are shifted by them. On the other hand, $L_{2+}$ causes the ``anomalous diffusion" that shifts only the covariance of the position and momentum. In this way we provide a different perspective in analyzing the damping modes of harmonic oscillator in open quantum systems.

\appendix

\section{Diffusion coefficients in the interaction picture}
\label{AppKd}

In this appendix, we obtain the diffusion coefficients of $\Kt_\dt$ \eqref{Kt}. Using the commutation relations in Table \ref{tab3}, we rewrite the commutator between $K_0$ and $K_\dt$ as products of matrices,
\begin{align}   \label{K0Kd}
        [K_0,K_\dt]&=\begin{array}{c}
                    (-\psi \eta_1 +\phi \eta_2)O_+\\
                    +(-\psi\eta_0+\th \eta_2)L_{1+}\\
                    +(\phi \eta_0-\th \eta_1) L_{2+}
                    \end{array}
                    \simeq\Kv_0\cdot \etav\,,
\end{align}
where $\etav$ and $\Kv_0$ are given by \Eqs{etav} and \eqref{K0v}, respectively. We use the symbol $\simeq$ to denote the fact that the first, the second and the third component of a column vector are the coefficients of $O_+, L_{1+}$ and $L_{2+}$, respectively. Similarly, $[K_0,[K_0,K_\dt]]\simeq\Kv_0^2\cdot \etav$ and so on. Therefore, $\etat_i(t)$ in \Eq{Kdt} can be written as $\etatv(t)=\exp(\Kv_0 t)\cdot\etav$.

To obtain \Eq{etatv}, we need to solve the non-hermitian eigenvalue problem $\Kv_0\cdot \uv=\lam \uv$ and $\vv^\dag\cdot\Kv_0=\lam \vv^\dag$, where $\uv$ and $\vv$ are the right and left eigenvectors of $\Kv_0$, respectively. The eigenvalues are $\lam_0=0, \lam_\pm=\pm i\w$, where $\w=\sqrt{\th^2-\phi^2-\psi^2}$ is assumed to be real. The corresponding zero eigenvectors $\uv_0, \vv_0$ are given in \Eq{uv}. The rests are
\begin{align}   \label{up}
    \uv_+&= \frac{1}{N}\left(\begin{array}{c}
                             \tth\phih+i\psih \\
                             1+\phih^2 \\
                             \phih\psih+i\tth
                           \end{array}\right)\,,&\qquad
    \vv_+&= \frac{1}{N}\left(\begin{array}{c}
                             -\tth\phih-i\psih \\
                             1+\phih^2 \\
                             \phih\psih+i\tth
                           \end{array}\right)\,,\\
    \uv_-&= -\uv_+^\dg \,,&\qquad
    \vv_-&=-\vv_+^\dg\,,\label{um}
\end{align}
where $N=\sqrt{2(1+\phih^2)}$ is a normalization factor. Furthermore, $\tth\equiv/\w, \phih\equiv \phi/\w$ and $\psih\equiv\psi/\w$. The eigenvectors obey the orthonormal relation $\vv_i^\dg\cdot \uv_j =\delta_{ij}$ for $i,j=0,\pm$.

Defining
\begin{align}   \label{UV}
    \Uv&\equiv \left(\begin{array}{ccc}
                             \uv_0 &\uv_+ &\uv_-
                           \end{array}\right)\,, &\qquad
    \Vv&\equiv \left(\begin{array}{ccc}
                             \vv_0 &\vv_+ &\vv_-
                           \end{array}\right)\,,
\end{align}
we can then show explicitly that $\Vv^\dg\cdot\Uv=\Iv$ and $\Uv\cdot\Vv^\dg=\Iv$.
Now $\Kv_0$ is diagonalized by
\begin{align}   \label{K0diag}
    \Vv^\dg\cdot\Kv_0\cdot\Uv=\text{diag}
                    \left( \begin{array}{ccc}
                             \lam_0 &\lam_+ &\lam_-
                           \end{array}\right)\,.
\end{align}
As a result,
\begin{align}   \label{etatvdiag}
        \etatv&=\Uv\cdot\Vv^\dg\cdot\exp(\Kv_0 t)\cdot\Uv\cdot\Vv^\dg\cdot \etav \no
            &=\Uv\cdot\exp(\Vv^\dg\cdot\Kv_0\cdot\Uv t)\cdot\Vv^\dg\cdot \etav \no
            &=\Uv\cdot  \left( \begin{array}{ccc}
                             1 &0 &0 \\
                             0 &e^{i \w t} &0 \\
                             0 &0 &e^{-i \w t}
                           \end{array}\right)
                           \cdot\Vv^\dg \cdot \etav\,.
\end{align}
Multiplying the matrices on right hand side gives us \Eqs{etatv}-\eqref{K0v}.

Substituting the coefficients of the master equations into \Eq{etatv}, we obtain the following results. For KL equation \eqref{KKL}, we find that
\begin{align} \label{KLetat}
       \etat_{0,\KL}(t)&=-2b\gam\,, &\qquad \etat_{1,\KL}(t)&=0=\etat_{2,\KL}(t)\,.
\end{align}
For CL equation \eqref{KCL}, we find that
\begin{subequations}
\begin{align} \label{CLetat}
       \etat_{0,\CL}(t)&=-2b\gam(1+\gamh^2+\gamh^2 \cos\w t+\gamh \sin\w t)\,, \\ \etat_{1,\CL}(t)&=-2b\gam(\cos \w t+\gamh \sin\w t)\,,\\
       \etat_{2,\CL}(t)&=2b\gam\tth\big(\gamh-\gamh\cos\w t+\sin \w t\big)\,,
\end{align}
\end{subequations}
where $\gamh\equiv \gam/\w$.
For HPZ equation \eqref{KHPZ}, we find that
\begin{subequations}
\begin{align} \label{HPZetat}
       \etat_{0,\HPZ}(t)&=\etat_{0,\CL}+d\tth\gamh ( \cos\w t-1)\,, \\ \etat_{1,\HPZ}(t)&=\etat_{1,\CL}-d\tth \sin\w t \,,\\
       \etat_{2,\HPZ}(t)&=\etat_{2,\CL}+d\big(\psih^2-(1+\psih^2)\cos\w t\big)\,.
\end{align}
\end{subequations}

\section{Commutation relations of generators}
\label{AppComm}

\begin{table}[ht]
\center
\begin{tabular}{c||ccc|cccc}
                & $iL_0$ & $iM_1$ & $iM_2$ & $O_0$ & $O_+$ & $L_{1+}$ &
                  $L_{2+}$  \\
                  \hline\hline
                 $iL_0$ & 0 & $-iM_2$ & $iM_1$ & 0 & 0 & $-L_{2+}$ & $L_{1+}$  \\
                 $iM_1$ & $iM_2$ & 0 & $iL_0$ & 0 &  $L_{2+}$ & 0 & $O_+$ \\
                 $iM_2$ &  $-iM_1$ &  $-iL_0$ & 0 & 0 &  $-L_{1+}$ &  $-O_+$ & 0  \\
                 \hline
                 $O_0$ & 0 & 0 & 0 & 0 &$O_+$ &$L_{1+}$ &$L_{2+}$ \\
                 $O_+$ & 0 &  $-L_{2+}$ &  $L_{1+}$ &  $-O_+$ & 0 & 0 & 0  \\
                 $L_{1+}$ &  $L_{2+}$ &  0 & $O_+$ & $-L_{1+}$ & 0 & 0 & 0 \\
                 $L_{2+}$ & $-L_{1+}$ & $-O_+$ & 0 & $-L_{2+}$ & 0 & 0 & 0 \\
\end{tabular}
\caption{Commutation relations of the generators.}
\label{tab3}
\end{table}

We reproduce the commutation relations of the generators \cite{Tay17} in Table \ref{tab3} for the convenience of the readers. The table is read as follows. For example, the commutation between $iM_2$ (in the first column) and $O_+$ (in the first row) gives $-L_{1+}$, i.e., $[iM_2, O_+]=-L_{1+}$, and etc.

\section{Generators in the phase space coordinates}
\label{AppGenxp}

To obtain the operators in the phase space coordinates, we need to first write the operators in the center and relative coordinate $(q,r)$, cf. Ref.~\cite{Tay17b}. Then we carry out a Fourier transform on the relative coordinate $r$ to obtain the representation of the operators in the phase space coordinate $(q,p)$, which we denote by the symbol $\doteq$,
\begin{subequations}
\begin{align} \label{iL0xp}
    iL_0 &\doteq -\frac{1}{2}\left(q\frac{\d}{\d p}-p\frac{\d}{\d q}\right)\,,&\quad
    iM_1 &\doteq -\frac{1}{2}\left(q\frac{\d}{\d p}+p\frac{\d}{\d q}\right)\,,\\
    iM_2 &\doteq -\frac{1}{2}\left(q\frac{\d}{\d q}-\frac{\d}{\d p}p+1\right)\,,&\quad
    O_0-I/2  &\doteq -\frac{1}{2}\left(q\frac{\d}{\d q}+\frac{\d}{\d p}p+1\right)\,,\\
    O_+  &\doteq \frac{1}{4}\left(\frac{\d^2}{\d q^2}+\frac{\d^2}{\d p^2}\right)\,,&\quad
   L_{1+} &\doteq \frac{1}{4}\left(-\frac{\d^2}{\d q^2}
                    +\frac{\d^2}{\d p^2}\right)\,,\\
    L_{2+} &\doteq \frac{1}{2}\frac{\d^2}{\d p \d q}\,.\label{L2+xp}
\end{align}
\end{subequations}


\providecommand{\noopsort}[1]{}\providecommand{\singleletter}[1]{#1}%

\end{document}